\documentclass[prb,twocolumn,superscriptaddress,showpacs]{revtex4}

\usepackage{graphicx}
\usepackage{mathrsfs}
\usepackage{amsmath,amssymb}

\begin{document}

\title{Comment on "Isotope effect in high-T$_c$ superconductors" (Physical Review B ${\bf 77}$, 024523 (2008))}

\author{A. S. Alexandrov}
\affiliation{Department of Physics, Loughborough University, Loughborough, LE11 3TU, U.K.}

\author{G. M. Zhao}
\affiliation{Department of Physics and Astronomy, California State University, Los Angeles, CA 90032, USA}

\begin{abstract}

We show that the recent reinterpretation of  oxygen isotope effects
in cuprate superconductors by D. R. Harshman \emph{et al.} is
mathematically and physically incorrect violating the Anderson
theorem and the Coulomb law.


\end{abstract}

\maketitle

The doping dependent oxygen isotope effect (OIE), $\alpha$,  on the
critical superconducting temperature $T_c$ (for recent reviews see
Ref.~\cite{polarons}) and the substantial OIE on the carrier mass
\cite{zhao}, $\alpha_{m^*}$,  provide  direct  evidence for a
significant electron-phonon interaction (EPI) in  cuprate
superconductors. High resolution angle-resolved photoemission
spectroscopy (ARPES) \cite{lanzara} provides further evidence for
the strong EPI  \cite{allen} apparently with c-axis-polarised
optical phonons. These results along with  optical
\cite{ita,mic,tal} and neutron scattering \cite{ega,rez}
spectroscopies unambiguously show that lattice vibrations  play a
significant but unconventional role in high-temperature
superconductivity. The interpretation of the optical spectra of
high-T$_c$ materials as due to many-polaron absorption
\cite{TDPRB01} strengthens the view \cite{asa} that the Fr\"{o}hlich
EPI is important in those structures. Operating together with a
short-range deformation potential and molecular-type (e.g.,
Jahn-Teller \cite{muller:2000}) EPIs, the Fr\"ohlich EPI can readily
overcome the Coulomb repulsion at a short distance of about the
lattice constant for electrons to form  real-space intersite
bipolarons or Cooper pairs depending on doping \cite{alemot}.

Despite  all these remarkable and well-done experiments that lead to
the consistent conclusion about the important role  of EPI in
high-temperature superconductors, Harshman \emph{et al.} \cite{har}
have recently claimed that the observed large OIE is caused by a
disorder-induced pair-breaking   rather than by strong
electron-phonon coupling and/or polaronic effects. Based on their
reinterpretation of OIE, they  conclude that EPI is allegedly too
weak to explain high $T_c$ in all the high-T$_c$ materials. Here we
show that the reinterpretation of OIE \cite{har} is internally
inconsistent being at odds with a couple of fundamental physical
laws. More specifically we show that the reinterpretation stems from
a mathematically incorrect formulism.

Given the added claim that the pairing symmetry is nodeless s-wave,the
authors in Ref.~\cite{har} have assumed that the variation of $T_c$ with doping is determined by the ``universal'' relation,
\begin{equation}
\ln (T_{co}/T_{c})=a[\psi (1/2 + 1/T_c \tau)-\psi(1/2)],\label{breaking}
\end{equation}
which was originally derived by Abrikosov and Gor'kov \cite{abr} with the coefficient $a=1$
to describe the pair-breaking effect by magnetic impurities in conventional s-wave BCS superconductors. Here $T_{c0}$ is the critical temperature of optimally doped compounds in
the absence of pair-breaking,  $\psi(x)$ is the digamma function, and  $\tau=4\pi \tau_{tr}$ is proportional to the
transport relaxation time, $\tau_{tr}$, due to impurities, which are thought to be responsible for the suppression of
$T_{c0}$ (we take $\hbar=k_B=1$ here and further). Since \emph{nonmagnetic} disorder in cuprate superconductors also often depresses
$T_{c0}$, Harshman \emph{et al.}\cite{har} have erroneously relaxed the requirement of magnetic impurities applying
Eq.~(\ref{breaking}) to nonmagnetic impurities with the same coefficient $a=1$.

In fact, the coefficient $a$ in Eq.(\ref{breaking})  strongly
depends on the pairing symmetry \cite{gorkov}  as analyzed in detail
by Fehrenbacher and Norman \cite{norman}. For nonmagnetic impurities
$a=1$ holds only for a d-wave (DW) or g-wave (GW) superconductor
with a zero average gap, while this coefficient is significantly
smaller in an anisotropic s-wave (ASW) superconductor \cite{norman}.
When the BCS gap is isotropic, the familiar ``Anderson theorem'',
$T_{c} = T_{c0}$, is satisfied \cite{and,abr2} because $a=0$. But
even in the extreme case of a highly anisotropic ASW superconductor
with the same nodal structure as in the DW superconductor the effect
of nonmagnetic impurities on their properties remains qualitatively
different, although the two states are indistinguishable in
phase-insensitive experiments \cite{norman}. In particular we show
here that  the pair-breaking OIE enhancement is negligibly small
based on any s-wave gap function that does not change sign with
angle, contrary to Ref.\cite{har}.

On the other hand, the effect of \emph{magnetic} impurities in an
ASW superconductor, or the effect of \emph{nonmagnetic} impurities
(or disorder) in a DW or GW superconductor can cause a significant
enhancement of the isotope effects on both $T_{c}$ and the
penetration depth \cite{Tallon1,Bill}. Two different groups
\cite{Tallon1,Bill} have consistently shown that the isotope effects
on  $T_{c}$ and the penetration depth are almost proportional to
each other provided that the strong pair-breaking effect exists.
These theoretical models may be able to explain the observed large
oxygen-isotope effects on both $T_{c}$ and the penetration depth in
underdoped cuprates if the scattering rate were large enough.
However, these models cannot consistently explain the negligibly
small OIE on $T_{c}$ but a large OIE on the penetration depth in
optimally doped cuprates \cite{zhao}.

Differentiating Eq.(\ref{breaking}) with respect to the ion mass, $M$, one can express OIE, $\alpha= -d \ln T_c/d \ln M$, in terms of the OIE
observed in optimal compounds, $\alpha_0= -d \ln T_{c0}/d \ln M$,
\begin{equation}
\alpha={\alpha_0\over{1-a\psi_1 (1/2 + 1/T_c \tau)/T_c
\tau}},\label{iso}
\end{equation}
where $\psi_1(x)=d\psi(x)/dx$ is the trigamma function, if  $\tau$ is independent of $M$. As shown in
Fig.~1 (right panel) using Eq.(\ref{iso}) the maximum OIE enhancement is about 30\% or less even in the extreme case of the ASW superconductor
with the same nodal structure as in the DW superconductor, where the enhancement is huge, about several hundred percent or more,
Fig.~1 (left panel). For a nodeless s-wave gap, hypothesized in Ref.~\cite{har}, there is practically no enhancement at all. Mathematically the difference
comes from the different numerical coefficients in Eq.(\ref{breaking}): $a=1$ for DW, $a=1/4$ for the extreme ASW
\cite{norman}, and $a < 1/4$ for a nodeless gap.
Physically the difference comes from the non-vanishing, impurity-induced, off-diagonal self energy in the ASW state, which is absent in
the DW state \cite{gorkov,norman}.
\begin{figure}
\includegraphics[width = 45mm, angle=-90]{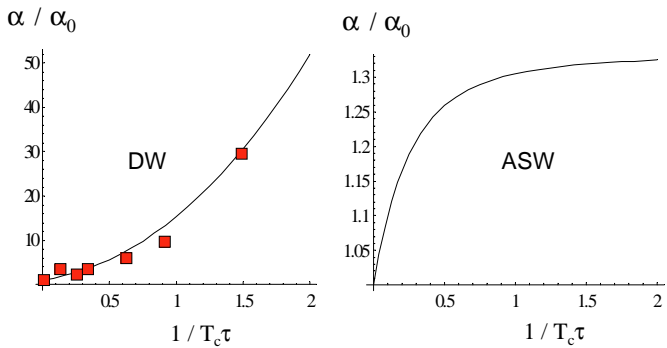}
\caption{Pair-breaking enhancement of the oxygen isotope effect,
$\alpha/\alpha_0$, in the d-wave (DW) superconductor (left panel)
and in an anisotropic s-wave (ASW) superconductor with the same
nodal structure (right panel) as a function of the pair-breaking
parameter, $1/\tau$. Symbols (right panel) represent the
experimental data for Zn-doped YBa$_2$Cu$_3$O$_{7-\delta}$ used in
Ref. \cite{har}.} \label{SWaveUInf}
\end{figure}
As a result the ``pair-breaking'' reinterpretation of OIE by Harshman
\emph{et al} \cite{har}  with the nodeless  pairing symmetry turns
out to be incompatible with the experimental data. The experimental
OIE, Fig.~1 (left panel), is  more than one order of magnitude larger
than the predicted OIE when the correct equation is applied, Fig.~1
(right panel).

One can also rule out the pair-breaking explanation of OIE
\cite{har} even in the case of the DW order parameter, in particular
in Pr substituted YBa$_{2}$Cu$_{3}$O$_{7-y}$ (YBCO), although there
is apparently good agreement with the experiment in the case of
Zn-doped YBCO, as seen in Fig.~1 (left panel). Since Zn doping
induces a magnetic moment of about 0.8 $\mu_{B}$ per Zn, the data
might be consistent with the magnetic pair-breaking effect in the
case of an s-wave symmetry. But for YBCO with oxygen vacancies or
substituted by trivalent elements for Ba, no magnetic moments and
disorder are induced in the CuO$_{2}$ planes so that the impurity
scattering rate may increase only slightly. In fact the
low-temperature coherence length in cuprate superconductors is very
small, $\xi_0=0.18 v_F/T_c < 2$nm, while the mean free path, $l=v_F
\tau_{tr}$, is about $10$ nm or larger as follows from resistivity
and recent quantum magneto-oscillation measurements in the
underdoped YBa$_{2}$Cu$_{3}$O$_{6.5}$ \cite{proust} ($v_F$ is the
Fermi velocity). Using these data one obtains $1/T_c\tau < 0.1$,
which is too small to account for the observed enhancement of OIE
with any gap symmetry as seen from Fig.~1, or for the doping
dependence of the magnetic penetration depth, contrary to
Ref.\cite{har}.

If however, in spite of the above estimate,  Pr substitution might
lead to a pair-breaking parameter, $1/T_c\tau \gtrsim 0.5$, large
enough to explain the enhancement of OIE on $T_{c}$, one should
expect a similar enhancement of  OIE on the penetration depth
because the magnitudes of the enhancement in the isotope effects on
$T_{c}$ and the penetration depth are nearly proportional to each
other \cite{Tallon1,Bill}.  Nevertheless,  OIE on the penetration
depth is nearly constant from the optimally doped sample to  the
substituted samples with a large amount of Pr \cite{har}, which is
inconsistent with the theoretical prediction \cite{Tallon1,Bill}.
Claiming the opposite, Harshman \emph{et al.}  have made further
mistakes in their derivation of the penetration depth,
$\lambda_{ab}$ (Eq.~(8) in Ref.\cite{har}). They have applied the
conventional correction factor  $1+\xi_0/l$ due to impurity
scattering,  which actually yields
\begin{equation}
\lambda_{ab}^2(T_c)=\lambda_{ab}^2(T_{co}) [1+ 0.36
\tilde{\alpha}/T_c],\label{correct}
\end{equation}
where $\tilde{\alpha}= 1/2\tau_{tr}$. Eq.(\ref{correct}) differs
from   Eq.~(8) in Ref.\cite{har} with $T_{c0}$ instead of $T_{c}$ in
the second term inside the square brackets. Clearly using
Eq.(\ref{correct}) instead of the incorrect Eq.~(8) in Ref.
\cite{har} one obtains an enhancement of OIE on $\lambda_{ab}$
similar to that on $T_{c}$ contrary to the erroneous claim of
Ref.\cite{har}. Moreover the penetration-depth formula
Eq.(\ref{correct}) is valid only for nonmagnetic impurities in
s-wave  superconductors that do not suppress $T_{c}$. If one assumes
that nonmagnetic impurities can suppress $T_{c}$, one should
consistently use a formula for the penetration depth, which is also
associated with the pair-breaking effect.

We would like to emphasize here that, since the pair-breaking effect
in optimally doped cuprates is negligibly small and the carrier
concentrations of the two oxygen-isotope samples  have been
consistently proved to be the same within $\pm$0.0002 per Cu
\cite{ZhaoJPCM,zhao}, the observed large oxygen-isotope effect on
the penetration depth must be caused by the large oxygen-isotope
effect on the supercarrier mass. The origin of this unconventional
isotope effect should arise from strong EPI that causes the
breakdown of the Migdal approximation. Indeed a model based on
(bi)polarons \cite{aleiso} accounts  naturally for both OIEs,
$\alpha$ and $\alpha_{m^*}$. There is a qualitative difference
between ordinary metallic and  polaronic conductors. The
renormalized effective mass of electrons is independent of the ion
mass $M$ in ordinary metals (where the Migdal adiabatic
approximation is believed to be valid), because the EPI constant
$\lambda=E_p/D$ does not depend on the isotope mass ($D$ is the
electron bandwidth in a rigid lattice). However, when electrons form
polarons dressed by lattice distortions, their effective mass
$m^{*}$ depends on $M$ through $m^{*}= m \exp (\gamma E_p/\omega)$,
where $m$ is the band mass in a rigid lattice and $\gamma <1$ is a
numerical coefficient depending on the EPI range. Here the phonon
frequency, $\omega$, depends on the ion mass, so that there is a
large polaronic isotope effect
 on the carrier mass with the carrier mass isotope exponent $\alpha_{m^*} = d\ln m^{*}/d\ln M=(1/2)\ln (m^*/m)$ as
observed \cite{zhao}, in contrast to the zero isotope effect in
ordinary metals. Importantly $\alpha_{m^*}$ is related to the
critical temperature isotope exponent, $\alpha$, of a (bi)polaronic
superconductor as $\alpha = \alpha_{m^*} [1 - (m/m^*)/(\lambda -
\mu_c)]$, where $\mu_c$ is the Coulomb pseudo-potential
\cite{aleiso}. Contrary to another misleading claim by Harshman
\emph{et al.} \cite{har}, the latter expression accounts for
different doping dependencies of $\alpha$ and $\alpha_{m^*}$ as well
as for a small value of $\alpha$ compared with $\alpha_{m^*}$ in
optimally doped samples, where the electron-phonon coupling constant
$\lambda$ approaches from above the Coulomb pseudopotential $\mu_c$
\cite{aleiso}. Similarly, the unconventional isotope effects
\cite{polarons} were also explained by polaron formation stemming
from the coupling to the particular quadrupolar $Q(2)$-type phonon
mode  in the framework of a multi-band polaron model
\cite{bussmanniso}.

Finally the claim by Harshman \emph{et al.} \cite{har} that EPI is
weak in high-T$_c$ superconductors compared with the Coulomb
coupling between  carriers in buffer and CuO$_{2}$ layers
contradicts  the Coulomb law.  EPI with $c-$axis polarized optical
phonons is virtually unscreened  since the upper limit for the
out-of-plane plasmon frequency  \cite{mar} ($\lesssim 200$
cm$^{-1}$) is well below the characteristic frequency of optical
phonons, $\omega \approx 400 \div 1000$ cm$^{-1}$ in all cuprate
superconductors. As the result of poor
 screening  the magnitude of an effective attraction  between carriers, $V_{ph}(r)=-e^2(\epsilon_{\infty}^{-1}-\epsilon_{0}^{-1})/r$, induced by the Fr\"ohlich EPI, is essentially the same as the  Coulomb repulsion,
 $V_{c}(r)=e^2/\epsilon_{\infty}r$, both of the order of $1$ eV, as directly
 confirmed by a huge difference in the static, $\epsilon_{0} \gg 1$,  and high-frequency, $\epsilon_{\infty}\thickapprox 4 \div 5$,  dielectric constants of these
 ionic crystals \cite{bra}.

To summarize we have shown that the conclusions by Harshman \emph{et
al.} \cite{har}  are mathematically erroneous and physically  at
odds with the fundamental Anderson theorem and the Coulomb law. We
thank  Annette Bussmann-Holder   for calling our attention to
Ref.\cite{har} and illuminating discussions.

 \bibliography{CommentDowF.bbl}

\end{document}